\documentclass{sig-alternate}

\usepackage{tikz}
\usepackage[all]{xy}
\usepackage{multirow}
\usepackage{amsmath}
\usepackage{amsfonts}
\usepackage{enumerate}
\usepackage{times}
\usepackage{bm}
\usepackage{algorithm}
\usepackage{algorithmic}
\usepackage{hyperref}

\AtBeginDocument{%
  \mathchardef\mathcomma\mathcode`\,
  \mathcode`\,="8000 
}
{\catcode`,=\active
  \gdef,{\mathcomma\discretionary{}{}{}}
}

\newcommand{\rulelabel}[1]{(\textsc{#1})}

\newcommand{\deriv}{\vdash_D}
\newcommand{\lub}{\sqcup}
\newcommand{\sub}{\sqsubset}

\newdef{definition}{Definition}
\newdef{lemma}{Lemma}
\newdef{corollary}{Corollary}
\newdef{theorem}{Theorem}
\newdef{example}{Example}
\newdef{remark}{Remark}
\newdef{proposition}{Proposition}
\newdef{property}{Property}
\newcommand{\tellp}[1]{\tell(#1)}
\newcommand{\askp}[2]{\ask \  #1 \  \rightarrow \ #2}

\newcommand{\defsymbol}{\stackrel{\textup{\texttt{def}}}  {=}}

\newcommand{\true}{{\it true}}
\newcommand{\false}{{\it false}}
\newcommand{\vars} {{\it Var}}
\newcommand{\ask}{{\bf ask}}

\newcommand{\tell}{{\bf tell}}
\newcommand{\Stop}{{\bf 0}}

\newcommand{\rrarrow}{\longrightarrow}

\newcommand{\Con}{{\it Con}}

\newcommand{\pairccp}[2]{\langle #1,#2 \rangle}
\newcommand{\trans}[1]{\stackrel{#1}{\rrarrow}}
\newcommand{\tr}[1]{\trans{#1}}
\newcommand{\localprocess}[3]{{\bm \exists}^{#2}_{#3} #1}
\newcommand{\truep}{\mbox{\true}}
\newcommand{\stopp}{\mbox{\Stop}}

\newcommand{\barbbis}{\dot{\sim}_b}
\newcommand{\satbis}{\dot{\sim}_{sb}}

\newcommand{\bigfrac}[2]{
\renewcommand{\arraystretch}{1.6}
\begin{array}{c}#1\\
\hline #2
\end{array}}

\renewcommand{\arraystretch}{1.5}
\long\def\comment#1{}

\makeatletter
\let\@copyrightspace\relax
\makeatother

\begin{document}

\title{Partition Refinement for Bisimilarity in CCP\titlenote{This work has been partially supported by the project ANR-09-BLAN-0169-01 PANDA,
and by the French Defence procurement agency (DGA) with two PhD grants.}}

\numberofauthors{4} 
\author{
\alignauthor
Andr\'es Aristiz\'abal \titlenote{Com\`{e}te, LIX, Laboratoire de l'\'{E}cole Polytechnique
associ\'e  \`a l'INRIA} \\
       \affaddr{CNRS/DGA}\\
       \email{andresaristi@ \\ lix.polytechnique.fr}
\alignauthor 
Filippo Bonchi \titlenote{ENS Lyon, Universit\'e de Lyon, LIP (UMR 5668 CNRS ENS Lyon UCBL INRIA), 46 All\'ee d'Italie, 69364 Lyon, France} \\
       \affaddr{CNRS}\\
       \email{filippo.bonchi@ens-lyon.fr}
\alignauthor Luis Fernando Pino \footnotemark[2] \\
       \affaddr{INRIA/DGA}\\
       \email{luis.pino@ \\ lix.polytechnique.fr}
\and  
\alignauthor Frank D. Valencia  \footnotemark[2]\\
       \affaddr{CNRS}\\
       \email{frank.valencia@ \\ lix.polytechnique.fr}
}

\maketitle
\begin{abstract}
Saraswat's concurrent constraint programming (ccp) is a mature formalism for 
modeling processes (or programs) that interact by telling and asking constraints in a global medium, called the store.
Bisimilarity is a standard behavioural equivalence in concurrency theory, but 
a well-behaved notion of bisimilarity for ccp has been proposed only recently.  
When the state space of a system is finite, the ordinary notion of bisimilarity 
can be computed via the well-known {\it partition refinement algorithm}, but unfortunately,
this algorithm does not work for ccp bisimilarity.

In this paper, we propose a variation of the partition refinement algorithm  for verifying ccp bisimilarity. To the best of our knowledge this is
the first work  providing for the automatic verification of program equivalence for ccp. 
\end{abstract}


\section{Introduction}

\emph{Bisimilarity} is the main representative of the so called behavioral equivalences, 
i.e., equivalence relations that determine when two processes (e.g., the specification and the implementation) behave in the same way.
Many efficient algorithms and tools for bisimilarity checking 
have been developed \cite{Victor:94:CAV,Fernandez:89:SCP,Ferrari:98:CAV}.
Among these, the \emph{partition refinement algorithm}
\cite{Kanellakis:83:PODC} is the best known: first it generates the
state space of a labeled transition system (LTS), i.e., the set of states reachable through the transitions; then,
it creates a partition equating all states and afterwards,
iteratively, refines these partitions by splitting non equivalent
states. At the end, the resulting partition equates all and only 
bisimilar states.

\emph{Concurrent Constraint Programming} (ccp)
\cite{Saraswat:90:POPL} is a formalism that
combines the traditional algebraic and operational view of process
calculi with a declarative one based upon first-order logic. In ccp,
processes (agents or programs) interact by \emph{adding} (or \emph{telling}) and \emph{asking} information
(namely, constraints) in a medium (\emph{store}).  

{\bf Problem}. The ccp formalism has been widely investigated and tested in terms of theoretical studies 
and the implementation of several ccp programming languages. From the \emph{applied computing point of view}, however, ccp lacks 
algorithms and tools to automatically verify program equivalence.  In this paper, we will give the first step towards automatic verification
of ccp program equivalences by providing an algorithm to automatically 
verify a ccp process (or program) equivalence from the literature. Namely, \emph{saturated barbed bisimilarity}. 



Saturated barbed bisimilarity ($\satbis$) for ccp was  introduced in \cite{Aristizabal:11:FOSSACS}. Two configurations are equivalent according to $\satbis$ if  (i) they have the same store, (ii) their transitions go into equivalent states and (iii) they are still 
equivalent when adding an arbitrary constraint to the store. In \cite{Aristizabal:11:FOSSACS}, a weak variant of $\satbis$
is shown to be \emph{fully abstract} w.r.t. the standard 
observational equivalence of \cite{Saraswat:91:POPL}.


Unfortunately, the standard partition refinement algorithm does not work for $\satbis$ because condition (iii)
requires to check all possible constraints that might be added to the store.
In this paper we introduce a modified partition refinement algorithm for $\satbis$.

We closely follow the approach in \cite{Bonchi:09:ESOP} that studies the notion of saturated bisimilarity from a more general 
perspective and proposes an abstract checking procedure. 

We first define a \emph{derivation relation $\deriv$} amongst the transitions of ccp processes:
$\gamma \tr{\alpha_1} \gamma_1 \deriv \gamma \tr{\alpha_2} \gamma_2$ which intuitively means that the latter transition
is a logical consequence of the former.

Then we introduce the notion of \emph{redundant transition}. Intuitively, a 
transition $\gamma\tr{\alpha_2}\gamma_2$ is redundant if there
exists another transition $\gamma\tr{\alpha_1}\gamma_1$
that logically implies it, that is $\gamma\tr{\alpha_1}\gamma_1 \deriv \gamma\tr{\alpha_2}\gamma_2'$ 
and $\gamma_2$ $\satbis$ $\gamma_2'$.
Now, if we consider the LTS having only non-redundant transitions,
the ordinary notion of bisimilarity coincides with $\satbis$. Thus, in principle, we could remove all the redundant
transitions and then check bisimilarity with the standard partition
refinement algorithm. But  how can we decide which transitions are
redundant, if redundancy itself depends on $\satbis$ ?

Our solution consists in computing $\satbis$ and redundancy
\emph{at the same time}. In the first step, the algorithm considers
all the states as equivalent and all the transitions (potentially redundant) as redundant. At any iteration, states are
discerned according to (the current estimation of) non-redundant
transitions and then non-redundant transitions are updated according
to the new computed partition.

A distinctive aspect of our algorithm is that in the initial
partition, we  insert not only the reachable states, but also
extra ones which are needed to check for redundancy. We prove that these additional states are finitely many 
and thus the termination of the algorithm is guaranteed whenever the original LTS is finite (as it is the case of the standard partition refinement).
Unfortunately, the number of these states might be exponential wrt the size of the original LTS, consequently
the worst-case running time is exponential.

{\bf Contributions.} We provide an algorithm that allows us to verify saturated barbed bisimilarity for ccp. To the best of our knowledge, this is the first  algorithm for the automatic verification of a ccp program equivalence. 
This is done in Sections \ref{sec:IRB} and \ref{sec:PR4CCP} by building upon the results of
\cite{Bonchi:09:ESOP}. In Section \ref{sec:term} and \ref{sec:comp}, we also show the termination and the complexity of the algorithm. We have implemented the algorithm in c++  and the code is available at \href{http://www.lix.polytechnique.fr/~andresaristi/strong/}{http://www.lix.polytechnique.fr/\textasciitilde andresaristi/strong/}.  

\section{Background} We now introduce the original standard partition refinement  \cite{Kanellakis:83:PODC} and concurrent constraint programming (ccp). 

\subsection*{Partition Refinement}\label{sec:PR} 

In this section we recall the partition refinement algorithm introduced in \cite{Kanellakis:83:PODC} for 
checking bisimilarity over the states of a \emph{labeled transition system} (LTS).  
Recall that an LTS can be intuitively seen as a graph where nodes represent states (of computation) 
and arcs represent transitions between states. A transition $P\tr{a}Q$ between $P$ and $Q$ labelled with $a$ can be 
typically thought of as  an evolution from $P$ to $Q$ provided that a condition $a$ is met. 

Let us now introduce some notation. Given a set $S$, a \emph{partition} of $S$ is a set of
\emph{blocks}, i.e., subsets of $S$, that are all disjoint and whose
union is $S$. We write $\{B_1\}\dots \{B_n\}$ to denote a partition consisting
of blocks $B_1, \dots, B_n$. A partition represents an equivalence relation
where equivalent elements belong to the same block. We write $P \mathcal{P} Q$ to mean that $P$ and $Q$ are equivalent in the partition $\mathcal{P}.$

The \emph{partition refinement algorithm} (see Alg. \ref{algo:satPR}) checks the bisimilarity of a set of initial states $IS$ as follows.
First, it computes $IS^{\star}$, that is the set of
all states that are reachable from $IS$. Then it creates the
partition $\mathcal{P}^0$ where all the elements of $IS^{\star}$ belong to
the same block (i.e., they are all equivalent). After the initialization, it iteratively refines the
partitions by employing the function $\mathbf{F}$, defined as follows: 
for all partitions $\mathcal{P}$,
$P \, \mathbf{F}(\mathcal{P})\, Q$ iff
\begin{itemize}
 \item if $P\tr{a}P'$ then exists $Q'$ s.t. $Q\tr{a}Q'$ and  $P'\, \mathcal{P} Q'$.
\end{itemize}
The algorithm terminates whenever two consecutive partitions are
equivalent. In such partition two states belong to the same block iff they are bisimilar.

Note that any iteration
splits blocks and never fuses them. For this reason if $IS^{\star}$
is finite, the algorithm terminates in at most $|IS^{\star}|$
iterations.

\begin{proposition}
If $IS^{\star}$ is finite, then the algorithm terminates and the
resulting partition equates all and only the bisimilar states.
\end{proposition}

\begin{algorithm}\caption{\texttt{Partition-Refinement($IS$)}}\label{algo:satPR}
\textbf{Initialization}
\begin{enumerate}
\item $IS^{\star}$ is the set of all processes reachable from $IS$,
\item $\mathcal{P}^0:=\{IS^{\star}\}$,
\end{enumerate}
\textbf{Iteration} $\mathcal{P}^{n+1}:=\mathbf{F}(\mathcal{P}^n)$,

\textbf{Termination} If $\mathcal{P}^n=\mathcal{P}^{n+1}$ then return $\mathcal{P}^n$.
\end{algorithm}

\subsection*{CCP}\label{sec:CCP}
We now recall the concurrent constraint programming process calculus (ccp) \cite{Saraswat:90:POPL,Saraswat:91:POPL}. In particular its notion of barbed saturated bisimilarity ($\satbis$)  \cite{Aristizabal:11:FOSSACS}. 



\emph{Constraint Systems.}\label{sec:constraintsystems}
The ccp model is parametric in a \emph{constraint system} specifying
the structure and interdependencies of the information that
processes can ask and tell. Following \cite{Saraswat:91:POPL,Boer:95:TCS}, we regard a
constraint system as a complete algebraic lattice structure.

\begin{definition}\label{def:constraintsystem}
A \emph{constraint system} {\bf C} is a complete algebraic lattice
$({\Con},{\Con}_0,\sqsubseteq,\sqcup,\true,\false)$ where ${\Con}$
(the set of constraints) is a partially ordered set w.r.t.
$\sqsubseteq$, ${\Con}_0$ is the subset of {\it finite} elements of
${\Con}$, $\sqcup$ is the lub operation, and $\true$, $\false$ are
the least and greatest elements of ${\Con}$, respectively.
\end{definition}

To capture local variables \cite{Saraswat:91:POPL} introduces cylindric constraint systems.
A \emph{cylindric constraint system} over an infinite set of variables $V$ is a constraint
system equipped with an operation $\exists_x$ for each $x \in V$. Broadly speaking $\exists_x$ has the properties of the existential quantification of $x$--e.g., $\exists_x c \sqsubseteq c, \exists_x\exists_y c = \exists_y\exists_x c$ and $\exists_x(c \sqcup \exists_x d)= \exists_x c \sqcup \exists_xd$.  For the sake of space, we do not formally introduce this notion as it is not crucial to our work--see \cite{Saraswat:91:POPL}.

Given a partial order $(C,\sqsubseteq)$, we say that $c$ is strictly
smaller than $d$ ($c\sqsubset d$) if $c\sqsubseteq d$ and
$c\neq d$. We say that $(C,\sqsubseteq)$ is \emph{well-founded} if
there exists no infinite descending chains $\dots \sqsubset c_n
\sqsubset \dots \sqsubset c_1 \sqsubset c_0$. For a set $A\subseteq
C$, we say that an element $m\in A$ is \emph{minimal} in $A$ if for
all $a\in A$, $a\not \sqsubset m$. We shall use $min(A)$ to denote
the set of all minimal elements of $A$. Well-founded order and
minimal elements are related by the following result.
\begin{lemma}\label{lemma:wellfounded}
Let $(C,\sqsubseteq)$ be a well-founded order and $A\subseteq C$. If
$a\in A$, then $\exists m \in min(A)$ s.t., $m\sqsubseteq a$.
\end{lemma}

\begin{remark} We shall assume that the constraint system is well-founded and, for practical reasons, that its $\sqsubseteq$ is decidable. 
\end{remark}

We now define the constraint system we  use in our examples. 

\begin{example} 
 Let ${\it Var}$ be a set of variables and $\omega$ be the set of natural numbers.
A variable assignment is a function $\mu: {\it Var} \longrightarrow
\omega$. We use $\mathcal{A}$ to denote the set of all
assignments, ${\mathcal P}(A)$ to denote the powerset of $\mathcal{A}$, $\emptyset$ the empty set and $\cap$ the intersection of sets.
Let us define the following constraint system: The set of constraints is ${\mathcal P}(\mathcal{A})$. We define $c \sqsubseteq d$  iff  $c \supseteq d$. The constraint $\false$ is $\emptyset$, while $\true$ is
$\mathcal{A}$. Given two constraints $c$ and $d$, $c \sqcup d$ is
the intersection $c\cap d$. By abusing the notation, we will often
use a formula like $x<n$ to denote the corresponding constraint, i.e.,
the set of all assignments that map $x$ in a number smaller than $n$. 
\end{example}

\begin{table*}
{\scriptsize
$$\makebox{R1} \quad \pairccp{\tellp{c}}{d} \trans{}
\pairccp{\Stop}{d \sqcup c} \qquad 
\makebox{R2} \quad \bigfrac{c \sqsubseteq d}{\pairccp{\askp{(c)}{P}}{d} \trans{} \pairccp{P}{d}} 
\qquad 
\makebox{R5} \quad \bigfrac{\pairccp{P}{e \sqcup \exists_x d} \trans{} \pairccp{P'}{e'
\sqcup \exists_x d}}
{\pairccp{\localprocess{P}{e}{x}}{d} \trans{} \pairccp{\localprocess{P'}{e'}{x}}{d \sqcup \exists_x e'}}
$$

$$
\makebox{R3} \quad \bigfrac{\pairccp{P}{d} \trans{}
\pairccp{P'}{d'}}{\pairccp{P \parallel Q}{d} \trans{}
\pairccp{P'\parallel Q}{d'}} \qquad 
\makebox{R4} \quad \bigfrac{\pairccp{P}{d} \trans{}
\pairccp{P'}{d'}}{\pairccp{P \, + \, Q}{d} \trans{}
\pairccp{P'}{d'}}
\qquad \makebox{R6} \quad \bigfrac{\pairccp{P[\vec{z}/\vec{x}]}{d} \trans{} \gamma'}
{\pairccp{p(\vec{z})}{d} \trans{} \gamma'} \text{ for } p(\vec{x}) \defsymbol P 
$$
}
\caption{Reduction semantics for ccp (the symmetric rules for R3 and R4 are omitted)} \label{tab:opersem}
\end{table*}
\emph{Syntax.}\label{sec:syntax} Let us presuppose a cylindric
constraint system ${\bf C}=
({\Con},{\Con}_0,\sqsubseteq,\sqcup,\true,\false)$ over a set of
variables $\vars$.  The ccp processes are given  by the following
syntax, 
$$ 
P,Q::= \Stop \mid \tell(c) \mid \ask(c)\rightarrow P \mid P
\parallel Q  \mid P
 +   Q \mid \localprocess{P}{c}{x} \mid p(\vec{z})
$$
where $c \in {\Con}_0, \, x \in \vars$, $\vec{z} \in \vars^*$.  

Intuitively, $\Stop$ represents termination, $\tell(c)$ adds the constraint (or partial information) $c$ to the store. 
The addition is performed regardless the generation of inconsistent information.  
The process $\ask(c)\rightarrow P$ may execute $P$ if $c$ is entailed from the
information in the store. The processes $P \parallel Q$ and $P \, + \,  Q$ stand, respectively, for the
{\it parallel execution} and {\it non-deterministic choice} of $P$ and $Q$;  
$\localprocess{}{c}{x}$ is a
{\it hiding operator}, namely it indicates that in
$\localprocess{P}{c}{x}$ the variable $x$ is {\it local} to $P$ and $c$ is some
local information (\emph{local store}) possibly containing $x$. 
A process $p(\vec{z})$ is said to be a
{\it procedure call} with identifier $p$ and actual parameters
$\vec{z}$. We presuppose that for each procedure call $p(z_1\ldots
z_m)$ there exists a unique  {\it procedure definition} possibly
\emph{recursive}, of the form $p(x_1\ldots x_m)\defsymbol P$ where
${\it fv}(P)\subseteq \{ x_1,\ldots , x_m \}$. 


\emph{Reduction Semantics.}\label{sec:operationalmodel} The operational semantics is given by transitions
between configurations. 
 A configuration is a pair $\pairccp{P}{d}$ representing a \emph{state} of
a system;  $d$ is a constraint representing the global store, and
$P$ is a process, i.e., a term of the syntax.  We use ${\it Conf}$ with
typical elements $\gamma,\gamma',\ldots$ to denote the set of
configurations.  The operational model of ccp is given by the transition relation $\rrarrow\;\subseteq{\it Conf}\times{\it
Conf}$ defined in Tab. \ref{tab:opersem}. Except for R5, these standard rules
are self-explanatory. We include R5 for completeness of the presentation but it is not necessary to understand
our work in the next section.  For the sake of space we refer the interested reader to \cite{Aristizabal:11:FOSSACS} for a detailed explanation of the rules.


\emph{Barbed Semantics.}\label{sec:barbsem} The authors in \cite{Aristizabal:11:FOSSACS} introduced a  barbed semantics for ccp.
Barbed equivalences have been introduced in \cite{Milner:92:ICALP}
for CCS, and  become the standard behavioural equivalences for
formalisms equipped with unlabeled reduction semantics. Intuitively,
\emph{barbs} are basic observations (predicates) on the states of a
system. 

In the case of ccp, barbs are taken from the underlying set $\Con_0$ of the constraint system.
A configuration $\gamma=\pairccp{P}{d}$ is said to \emph{satisfy} the barb $c$
($\gamma \downarrow_c$) iff $c \sqsubseteq d$. 
%
\begin{definition}\label{def:barbbis}
A \emph{barbed bisimulation} is a symmetric relation $\mathcal{R}$ on
configurations s.t. whenever $ (\gamma_1,\gamma_2) \in
{\mathcal{R}} $:
\begin{enumerate}[(i)]
\item if $\gamma_1 \downarrow_c$ then $\gamma_2 \downarrow_c$,
\item if $\gamma_1 \trans{} \gamma_1'$ then there exists
$\gamma_2'$ s.t. $\gamma_2 \trans{} \gamma_2'$ and $(\gamma_1',
\gamma_2') \in {\mathcal{R}}$.
\end{enumerate}
$\gamma_1$ and $\gamma_2$ are \emph{barbed bisimilar} ($\gamma_1 \;  \barbbis \; \gamma_2$), if there exists a barbed
bisimulation $\mathcal{R}$ s.t. $(\gamma_1,\gamma_2) \in
\mathcal{R}$. 
\end{definition}

One can verify that $\barbbis$ is an equivalence.
However, it is not a \emph{congruence}; i.e., it is not preserved
under arbitrary contexts (the interested reader can check Ex. 7 in \cite{Aristizabal:11:FOSSACS}).
An elegant solution to modify bisimilarity  for obtaining a
congruence consists in \emph{saturated bisimilarity} \cite{Bonchi:06:LICS, Bonchi:09:FOSSACS} (pioneered by \cite{Montanari:92:FI}). 
The basic idea is simple: saturated bisimulations  are closed w.r.t. all
the possible contexts of the language. In the case of ccp, it is
enough to require that bisimulations are \emph{upward closed} as in condition $(iii)$ below.

\begin{definition}\label{def:satbarbbis}
A \emph{saturated barbed bisimulation} is a symmetric relation
$\mathcal{R}$ on configurations s.t. whenever
$(\gamma_1,\gamma_2) \in {\mathcal{R}} $  with $\gamma_1=
\pairccp{P}{d}$ and $\gamma_2=  \pairccp{Q}{e}$:
\begin{enumerate}[(i)]
\item if $\gamma_1 \downarrow_c$ then $\gamma_2 \downarrow_c$,
\item if $\gamma_1 \trans{} \gamma_1'$ then there exists
$\gamma_2'$ s.t. $\gamma_2 \trans{} \gamma_2'$ and $(\gamma_1',
\gamma_2') \in {\mathcal{R}}$,
\item for every $a\in \Con_0$, $(\pairccp{P}{d \sqcup a}, \pairccp{Q}{e \sqcup a}) \in \mathcal{R}$.
\end{enumerate}
$\gamma_1$ and $\gamma_2$ are \emph{saturated barbed
bisimilar} ($\gamma_1  \; \satbis \; \gamma_2$) if there
exists a saturated barbed bisimulation $\mathcal{R}$ s.t.
$(\gamma_1,\gamma_2) \in \mathcal{R}$. 
\end{definition}

\begin{example} \label{ex:satbarbbis1}
Take $T=\tellp{\true}$, $P=\askp{(x<7)}{T}$ and $Q=\askp{(x<5)}{T}$. You can see that  
$\pairccp{P}{true} \not \!\! \satbis \pairccp{Q}{true}$, since $\pairccp{P}{x<7} \tr{}$, while
$\pairccp{Q}{x<7} \not \! \! \tr{}$. Consider now the configuration $\pairccp{P+Q}{true}$ and 
observe that $\pairccp{P+Q}{true} \satbis \pairccp{P}{true}$. Indeed, for all constraints $e$, s.t.
$x<7 \sqsubseteq e$, both the configurations evolve into $\pairccp{T}{e}$, while for all $e$ s.t.
$x<7 \not \sqsubseteq e$, both configurations cannot proceed. Since $x<7 \sqsubseteq x<5$, 
the behaviour of $Q$ is somehow absorbed by the behaviour of $P$.
\end{example}
\begin{example} \label{ex:satbarbbis2}
Since $\satbis$ is upward closed, $\pairccp{P+Q}{z<5} \satbis \pairccp{P}{z<5}$
follows immediately by the previous example. Now take $R=\askp{(z<5)}{(P+Q)}$ and $S=\askp{(z<7)}{P}$. By analogous arguments of the previous example, one can show that
$\pairccp{R+S}{true} \satbis \pairccp{S}{true}\text{.}$
\end{example}
\begin{example}\label{ex:satbarbbis3}
Take $T'=\tellp{y=1}$, $Q'=\askp{(x<5)}{T'}$ and $R'=\askp{(z<5)}{P+Q'}$.
Observe that $\pairccp{P+Q'}{z<5} \not  \! \! \satbis \pairccp{P}{z<5}$ 
and that $\pairccp{R'+S}{true} \not \! \! \satbis \pairccp{S}{true}$, since 
$\pairccp{P+Q'}{x<5}$ and $\pairccp{R'+S}{true}$ can reach a store  containing the constraint $y=1$.
\end{example}
In \cite{Aristizabal:11:FOSSACS}, a {\it weak variant} 
of $\satbis$ is introduced and it is shown that it is {\it fully abstract} w.r.t. the 
standard observational equivalence of \cite{Saraswat:91:POPL}. In this paper, 
we will show an algorithm for checking $\satbis$ and we leave, as future work, to extend it for the weak semantics. 

Nevertheless, the equivalence $\satbis$ would seem hard to (automatically) check because of the upward-closure (namely, the
quantification over all possible $a\in \Con_0$ in condition $(iii)$)
of Def. \ref{def:satbarbbis}. The work in \cite{Aristizabal:11:FOSSACS} deals with this issue by refining the notion of
transition by adding to it a {\it label} that carries additional information about the constraints that cause the reduction.

\begin{table*}[t!]
{\scriptsize
$$
\makebox{LR1}\pairccp{\tellp{c}}{d} \trans{\true}
\pairccp{\Stop}{d \sqcup c} \quad 
\makebox{LR2}\bigfrac{\alpha \in \min \{a\in \Con_0 \, | \, c
\sqsubseteq d \sqcup a \ \}} {\pairccp{\askp{(c)}{P}}{d}
\trans{\alpha} \pairccp{P}{d \sqcup \alpha}} \quad
\makebox{LR3}
\bigfrac{\pairccp{P}{d} \trans{\alpha} \pairccp{P'}{d'}}
{\pairccp{P\parallel Q}{d} \trans{\alpha} \pairccp{P'\parallel
Q}{d'}} \quad
\makebox{LR4}
\bigfrac{\pairccp{P}{d} \trans{\alpha} \pairccp{P'}{d'}}
{\pairccp{P + Q}{d} \trans{\alpha} \pairccp{P'}{d'}} 
$$

$$
\makebox{LR5}\bigfrac{\pairccp{P[z/x]}{e[z/x] \sqcup d}
\trans{\alpha} \pairccp{P'}{e' \sqcup d \sqcup \alpha}}
{\pairccp{\localprocess{P}{e}{x}}{d} \trans{\alpha}
\pairccp{\localprocess{P'[x/z]}{e'[x/z]}{x}}{\exists_x (e'[x/z])
\sqcup d \sqcup \alpha }} \\~ \ {x \not\in {\it fv}(e'), z \not\in
{\it fv}(P) \atop \cup {\it fv}(e\sqcup d \sqcup \alpha)}
\makebox{LR6}
\bigfrac{\pairccp{P[\vec{z}/\vec{x}]}{d} \trans{\alpha} \gamma'}
{\pairccp{p(\vec{z})}{d} \trans{\alpha} \gamma'} \quad
\mbox{for } p(\vec{x}) \defsymbol P 
$$
}
\caption{Labeled semantics for ccp. (the symmetric rules for LR3 and LR4 are omitted)}\label{tab:labsem} 
\end{table*}

\emph{Labeled Semantics.}\label{sec:labsem} As explained in \cite{Aristizabal:11:FOSSACS}, in a transition of the form
$\pairccp{P}{d} \trans{\alpha} \pairccp{P'}{d'}$ the label $\alpha$
represents a \emph{minimal} information (from the environment)
that needs to be added to the store $d$ to evolve from  $\pairccp{P}{d}$ into $\pairccp{P'}{d'}$, i.e.,
$\pairccp{P}{d \sqcup \alpha} \trans{} \pairccp{P'}{d'}$.
The labeled transition relation $\rrarrow\;\subseteq{\it Conf}\times \Con_0
\times{\it Conf}$ is defined by the rules in Tab. \ref{sec:labsem}. 
The rule LR2, for example, says that $\pairccp{\askp{(c)}{P}}{d}$ can evolve to
$\pairccp{P}{d\sqcup \alpha}$ if the environment provides a minimal
constraint $\alpha$ that added to the store $d$ entails $c$, i.e.,
$\alpha \in \min \{a \in \Con_0 \, | \, c \sqsubseteq d \sqcup a
\}$. Note that assuming that $(\Con, \sqsubseteq)$ is well-founded
(Sec. \ref{sec:constraintsystems}) is necessary to guarantee that $\alpha$ exists
whenever $ \{a \in \Con_0 \, | \, c \sqsubseteq d \sqcup a \ \}$ is
not empty.  The other rules, except LR4, are easily seen to realize the above intuition. An explanation of LR5 is not needed to understand the present work. For the sake of space, we refer the reader to \cite{Aristizabal:11:FOSSACS} for a more detailed explanation of these labeled rules.
%
Fig. \ref{fig:ltsexample} illustrates the LTSs of our running example.

\begin{figure*}
\begin{center}
{\small
\begin{tikzpicture}
\node (defT) at (-1.1,4.2) {{\footnotesize $T = \tellp{\truep}$}};
\node (defT') at (-1,3.8) {{\footnotesize $T' = \tellp{y=1}$}};
\node (defP) at (2,4.2) {{\footnotesize $P = \askp{(x<7)}{T}$}};
\node (defS) at (2,3.8) {{\footnotesize $S = \askp{(z<7)}{P}$}};
\node (defQ) at (5.5,4.2) {{\footnotesize $Q = \askp{(x<5)}{T}$}};
\node (defQ') at (5.6,3.8) {{\footnotesize $Q' = \askp{(x<5)}{T'}$}};
\node (defR) at (9.5,4.2) {{\footnotesize $R = \askp{(z<5)}{(P+Q)}$}};
\node (defR') at (9.6,3.8) {{\footnotesize $R' = \askp{(z<5)}{(P+Q')}$}};
\node (RS) at (-1,1) {$\pairccp{R+S}{\truep}$};
\node (S) at (-1,2) {$\pairccp{S}{\truep}$};
\node (R'S) at (-1,3) {$\pairccp{R'+S}{\truep}$};
\node (PQ') at (3,3) {$\pairccp{P+Q'}{z<5}$};
\node (P) at (3,2) {$\pairccp{P}{z<7}$};
\node (PQ) at (3,1) {$\pairccp{P+Q}{z<5}$};
\node (P2) at (3,0) {$\pairccp{P}{z<5}$};
\node (T') at (6.5,3) {$\pairccp{T'}{z<5 \sqcup x<5}$};
\node (T1) at (6.5,2) {$\pairccp{T}{z<7 \sqcup x<7}$};
\node (T2) at (6.5,1) {$\pairccp{T}{z<5 \sqcup x<5}$};
\node (T3) at (6.5,0) {$\pairccp{T}{z<5 \sqcup x<7}$};
\node (S1) at (11,3) {$\pairccp{\stopp}{z<5 \sqcup x<5 \sqcup y=1}$};
\node (S2) at (11,2) {$\pairccp{\stopp}{z<7 \sqcup x<7}$};
\node (S3) at (11,1) {$\pairccp{\stopp}{z<5 \sqcup x<5}$};
\node (S4) at (11,0) {$\pairccp{\stopp}{z<5 \sqcup x<7}$};
\draw (PQ') to [out=90, in=0] (2.5,3.5) -- (-2.2,3.5) -- (-2.2,-0.5) -- node[above] {{\scriptsize $x<7$}} (6,-0.5) to [out=0,in=270,->] (T3);
\draw (R'S) to [->] node[above] {{\scriptsize $z<5$}} (PQ');
\draw (R'S) to [->] node[above] {{\scriptsize $z<7$}} (P);
\draw (S) to [->] node[above] {{\scriptsize $z<7$}} (P);
\draw (RS) to [->] node[above] {{\scriptsize $z<5$}} (PQ);
\draw (RS) to [->] node[above] {{\scriptsize $z<7$}} (P);
\draw (PQ') to [->] node[above] {{\scriptsize $x<5$}} (T');
\draw (P) to [->] node[above] {{\scriptsize $x<7$}} (T1);
\draw (PQ) to [->] node[above] {{\scriptsize $x<5$}} (T2);
\draw (PQ) to [->] node[above] {{\scriptsize $x<7$}} (T3);
\draw (P2) to [->] node[above] {{\scriptsize $x<7$}} (T3);
\draw (T') to [->] node[above] {{\scriptsize $\truep$}} (S1);
\draw (T1) to [->] node[above] {{\scriptsize $\truep$}} (S2);
\draw (T2) to [->] node[above] {{\scriptsize $\truep$}} (S3);
\draw (T3) to [->] node[above] {{\scriptsize $\truep$}} (S4);
\end{tikzpicture}}
\end{center}
\caption{The labeled transition systems of the running example ($IS = \{\pairccp{R'+S}{\true},\pairccp{S}{\true},\pairccp{R+S}{\true}\}$).}\label{fig:ltsexample} 
\end{figure*}

\begin{figure*}
\begin{tiny}
$
\begin{array}{lcl}
\mathcal{P}^0 & = & \{\pairccp{R'+S}{\true}, \pairccp{S}{\true}, \pairccp{R+S}{\true}  \}, \{\pairccp{P+Q'}{z<5},\pairccp{P+Q}{z<5},\pairccp{P}{z<5} \}, \{ \pairccp{P}{z<7}\}, \{\pairccp{T'}{z<5 \sqcup x<5},\pairccp{T}{z<5 \sqcup x<5},\\
&& \pairccp{\Stop}{z<5 \sqcup x<5}\},\{\pairccp{T}{z<7 \sqcup x<7}, \pairccp{\Stop}{z<7 \sqcup x<7}\}, \{\pairccp{T}{z<5 \sqcup x<7},\pairccp{\Stop}{z<5 \sqcup x<7}\},\{\pairccp{\Stop}{z<5 \sqcup x<5 \sqcup y=1}\}\\
\mathcal{P}^1 & = & \{\pairccp{R'+S}{\true}, \pairccp{S}{\true}, \pairccp{R+S}{\true}  \}, \{\pairccp{P+Q'}{z<5},\pairccp{P+Q}{z<5},\pairccp{P}{z<5} \}, \{ \pairccp{P}{z<7}\}, \{\pairccp{T'}{z<5 \sqcup x<5}\},\{\pairccp{T}{z<5 \sqcup x<5}\},\\
&& \{\pairccp{\Stop}{z<5 \sqcup x<5}\},\{\pairccp{T}{z<7 \sqcup x<7}\},\{ \pairccp{\Stop}{z<7 \sqcup x<7}\}, \{\pairccp{T}{z<5 \sqcup x<7}\},\{\pairccp{\Stop}{z<5 \sqcup x<7}\},\{\pairccp{\Stop}{z<5 \sqcup x<5 \sqcup y=1}\}\\
\mathcal{P}^2 & = & \{\pairccp{R'+S}{\true}, \pairccp{S}{\true}, \pairccp{R+S}{\true}  \}, \{\pairccp{P+Q'}{z<5}\},\{\pairccp{P+Q}{z<5},\pairccp{P}{z<5} \}, \{ \pairccp{P}{z<7}\}, \{\pairccp{T'}{z<5 \sqcup x<5}\},\{\pairccp{T}{z<5 \sqcup x<5}\},\\
&& \{\pairccp{\Stop}{z<5 \sqcup x<5}\},\{\pairccp{T}{z<7 \sqcup x<7}\},\{ \pairccp{\Stop}{z<7 \sqcup x<7}\}, \{\pairccp{T}{z<5 \sqcup x<7}\},\{\pairccp{\Stop}{z<5 \sqcup x<7}\},\{\pairccp{\Stop}{z<5 \sqcup x<5 \sqcup y=1}\}\\
\mathcal{P}^3 & = & \{\pairccp{R'+S}{\true}\},\{ \pairccp{S}{\true}, \pairccp{R+S}{\true}  \}, \{\pairccp{P+Q'}{z<5}\},\{\pairccp{P+Q}{z<5},\pairccp{P}{z<5} \}, \{ \pairccp{P}{z<7}\}, \{\pairccp{T'}{z<5 \sqcup x<5}\},\{\pairccp{T}{z<5 \sqcup x<5}\},\\
&& \{\pairccp{\Stop}{z<5 \sqcup x<5}\},\{\pairccp{T}{z<7 \sqcup x<7}\},\{ \pairccp{\Stop}{z<7 \sqcup x<7}\}, \{\pairccp{T}{z<5 \sqcup x<7}\},\{\pairccp{\Stop}{z<5 \sqcup x<7}\},\{\pairccp{\Stop}{z<5 \sqcup x<5 \sqcup y=1}\}\\
\mathcal{P}^4 & = & \mathcal{P}^3
\end{array}
$
\end{tiny}
\caption{The partitions computed by \texttt{CCP-Partition-Refinement($\{\pairccp{R'+S}{\true},\pairccp{S}{\true},\pairccp{R+S}{\true}\}$)}.}\label{fig:part} 
\end{figure*}

\emph{Syntactic Bisimilarity.} When defining bisimilarity over a LTS, barbs
are not usually needed because they can be somehow inferred from the
labels of the transitions. For instance, in CCS,
$P\downarrow_{a}$ iff $P\trans{a}$.  However this is not the case of ccp: barbs cannot be removed from the
definition of bisimilarity because they cannot be inferred from the
transitions. 

Taking into account the barbs, the obvious adaptation of labeled bisimilarity for ccp is the following:

\begin{definition}\label{def:synbis}\cite{Aristizabal:11:FOSSACS}
A \emph{syntactic bisimulation} is a symmetric relation $\mathcal{R}$ on
configurations s.t. whenever $(\gamma_1,\gamma_2)\in
\mathcal{R}$:
\begin{enumerate}[(i)]
\item  if $\gamma_1 \downarrow_c$ then $\gamma_2
\downarrow_c$,
\item if $\gamma_1 \trans{\alpha} \gamma_1'$ then
$\exists \gamma_2'$ s.t. $\gamma_2 \trans{\alpha} \gamma_2'$
and $(\gamma_1',\gamma_2') \in \mathcal{R}$.
\end{enumerate}
$\gamma_1$ and $\gamma_2$ are syntactically bisimilar,
($\gamma_1 \sim_S \gamma_2$) if there exists a
syntactic bisimulation $\mathcal{R}$ s.t. $(\gamma_1, \gamma_2)
\in \mathcal{R}$. 
\end{definition}

Unfortunately as shown in \cite{Aristizabal:11:FOSSACS} $\sim_S$ is  over-discriminating. 
As an example, consider the configurations 
$\pairccp{P+Q}{z<5}$ and
$\pairccp{P}{z<5} $, whose LTS is shown in Fig. \ref{fig:ltsexample}. They are not equivalent
according to $\sim_S$. Indeed $\pairccp{P+Q}{z<5} \trans{x<5}$, while $\pairccp{P}{z<5} \not \trans{x<5}$.
However they are equivalent according to $\satbis$ (Ex. \ref{ex:satbarbbis2}).


\section{Irredundant Bisimilarity}\label{sec:IRB}

Syntactic bisimilarity is over-discriminating because of some {\it redundant transitions}. For instance, 
consider the transitions: \\ (a) $\pairccp{P+Q}{z<5} \trans{x<7} \pairccp{T}{z<5 \sqcup x<7}$; \\ (b) $\pairccp{P+Q}{z<5} \trans{x<5} \pairccp{T}{z<5 \sqcup x<5}$.
\\ Transition (a) means that 
for all constraints $e$ s.t. $x<7 \sqsubseteq e$, \\
(c)$\pairccp{P+Q}{z<5\sqcup e} \trans{} \pairccp{T}{z<5 \sqcup e}$,
while transition (b) means that the reduction (c) is possible for all $e$ s.t. $x<5 \sqsubseteq e$.
Since $x<7 \sqsubseteq x<5$, transition (b) is ``redundant'', in the sense that its meaning is ``logically derived'' by transition (a). 

The following notion captures the above intuition:

\begin{definition}\label{def:deriv}
We say that $\pairccp{P}{c} \trans{\alpha} \pairccp{P_1}{c'}$ derives $\pairccp{P}{c} \trans{\beta} \pairccp{P_1}{c''}$, written 
$\pairccp{P}{c} \trans{\alpha} \pairccp{P_1}{c'} \deriv \pairccp{P}{c} \trans{\beta} \pairccp{P_1}{c''}$, 
iff there exists $e$ s.t. the following conditions hold: \\
\begin{tabular}{ccc}
 (i) $\beta = \alpha \lub e$ &
 (ii) $c'' = c' \lub e$ &
 (iii) $\alpha \neq \beta$
\end{tabular}
\end{definition}

One can verify in the above example that (a) $\deriv$ (b). Notice that in order to check if $\pairccp{P+Q}{z<5} \satbis \pairccp{P}{z<5}$, 
we could first remove the redundant transition (b) and then check $\sim_S$.

More generally, a naive approach to compute $\satbis$ would be to first remove all those transitions that can be derived by others, 
and then apply the partition refinement algorithm. However, this approach would fail since it would distinguish $\pairccp{R+S}{true}$ 
and $\pairccp{S}{true}$ that, instead, are in $\satbis$ (Ex. \ref{ex:satbarbbis2}). Indeed, $ \pairccp{R+S}{true}$ can perform: \\
(e) $\pairccp{R+S}{true} \trans{z<7} \pairccp{P}{z<7}$, \\
(f) $ \pairccp{R+S}{true} \trans{z<5} \pairccp{P+Q}{z<5}$, \\
while $\pairccp{S}{true} \not \trans{z<5}$. 
Note that transition (f) cannot be derived by other transitions, since (e) $\not \deriv$ (f).
Indeed, $P$ is syntactically different from $P+Q$, even if they have the same behaviour when inserted in the store $z<5$, i.e., 
$\pairccp{P}{z<5}\satbis \pairccp{P+Q}{z<5}$ (Ex. \ref{ex:satbarbbis2}). The transition (f) 
is also ``redundant'', since its behaviour ``does not add anything'' to the behaviour of (e).

\begin{definition}\label{def:redtran}
Let $\mathcal{R}$ be a relation and $\gamma \stackrel{\alpha}{\rightarrow} \gamma_1$ 
and $\gamma \stackrel{\beta}{\rightarrow} \gamma_2$ be two transitions. 
We say that the former dominates the latter one in $\mathcal{R}$ 
(written $\gamma \stackrel{\alpha}{\rightarrow} \gamma_1 \succ_\mathcal{R} \gamma \stackrel{\beta}{\rightarrow} \gamma_2$) 
iff 
\\ \begin{tabular}{ccc}
 (i) $\gamma \stackrel{\alpha}{\rightarrow} \gamma_1 \vdash_D \gamma \stackrel{\beta}{\rightarrow} \gamma_2'$  &
 (ii) $(\gamma_2', \gamma_2) \in \mathcal{R}$ &
\end{tabular} \\
A transition is redundant w.r.t. $\mathcal{R}$ if it is dominated in $\mathcal{R}$ by another transition. Otherwise, it is irredundant.
\end{definition}
Note that the transition $\gamma \stackrel{\beta}{\rightarrow} \gamma_2'$ might not be generated by the rules 
in Tab. \ref{tab:labsem}, but simply derived by $\gamma \stackrel{\alpha}{\rightarrow} \gamma_1$ through $\deriv$. 
For instance, transition (e) dominates (f) in $\satbis$, because 
(e) $\deriv  \pairccp{R+S}{true} \trans{z<5} \pairccp{P}{z<5}$ and $\pairccp{P}{z<5} \satbis \pairccp{P+Q}{z<5}$. 

Therefore, we could compute $\satbis$, by removing all those transitions that are redundant wrt $\satbis$.  This, however, would lead us 
to a circular situation: How to decide which transitions are redundant when redundancy itself depends on $\satbis$.

Our solution relies on the following definition that allows to compute bisimilarity and redundancy \emph{at the same time}.
\begin{definition}\label{def:irbis}
An irredundant bisimulation is a symmetric relation $\mathcal{R}$ on
configurations s.t. whenever $(\gamma_1,\gamma_2) \in
{\mathcal{R}} $:
\begin{enumerate}[(i)]
\item  if $\gamma_1 \downarrow_c$ then $\gamma_2
\downarrow_c$,
\item if $\gamma_1 \trans{\alpha} \gamma_1'$ is irredundant in $\mathcal{R}$ then
$\exists \gamma_2'$ s.t. $\gamma_2 \trans{\alpha}
\gamma_2'$ and $(\gamma_1',\gamma_2') \in \mathcal{R}$.
\end{enumerate}
$\gamma_1$ and $\gamma_2$ are irredundant bisimilar ($\gamma_1 \sim_I \gamma_2$), 
if there exists an irredundant bisimulation $\mathcal{R}$ s.t. $(\gamma_1,\gamma_2) \, \in \, \mathcal{R}.$
\end{definition}

\begin{theorem}\label{theo:coincidence}
$\sim_I=\satbis$
\end{theorem}
\begin{proof}
See \cite{sac2012-extended-version}.
\end{proof}

\section{Partition Refinement for CCP}\label{sec:PR4CCP}
Recall that we mentioned in Sec. \ref{sec:barbsem} that checking $\satbis$ seems hard because of the quantification
over all possible constraints.  However, by using Theo. \ref{theo:coincidence} we shall introduce an algorithm for checking $\satbis$  by employing the notion of irredundant bisimulation. 

The first novelty w.r.t. the standard partition refinement (Alg. \ref{algo:satPR}) consists in using {\it barbs}. Since configurations satisfying different barbs are 
surely different, we can safely start with a partition that equates all and only those states 
satisfying the same barbs. Note that two configurations satisfy the same barbs iff they have the same store. 
Thus, we take as initial partition 
$\mathcal{P}^0=\{IS^{\star}_{d_1}\}\dots \{IS^{\star}_{d_n}\}$, where $IS^{\star}_{d_i}$ is the subset of 
the configurations of $IS^{\star}$ with store $d_i$.

Another difference is that instead of using the function $\mathbf{F}$ of Alg. \ref{algo:satPR}, 
we refine the partitions by employing the function $\mathbf{IR}$ defined as follows:
for all partitions $\mathcal{P}$,
$\gamma_1 \, \mathbf{IR}(\mathcal{P})\, \gamma_2$ iff
\begin{itemize}
 \item if $\gamma_1\tr{\alpha}\gamma_1'$ is irredundant in $\mathcal{P}$, then there exists $\gamma_2'$ s.t. $\gamma_2\tr{\alpha}\gamma_2'$ and  $\gamma_1'\, \mathcal{P} \gamma_2'$.
\end{itemize}
It is now important to observe that in the computation of $\mathbf{IR}(\mathcal{P}^n)$, there might be involved also states that are not 
reachable from the initial states $IS$. For instance, consider the LTSs of $\pairccp{S}{\true}$ and $\pairccp{R+S}{\true}$ in Fig. \ref{fig:ltsexample}. 
The state $\pairccp{P}{z<5}$ is not reachable but is needed to check if $\pairccp{R+S}{\true} \tr{z<5} \pairccp{P+Q}{z<5}$ is redundant (look at the example after Def. \ref{def:redtran}).

For this reason, we have also to change the initialization step of
our algorithm, by including in the set $IS^{\star}$ all the states
that are needed to check redundancy. This is done, by using the
following closure rules.
{\scriptsize
$$\rulelabel{is}\bigfrac{\gamma \in IS}{\gamma \in IS^{\star}} \qquad \qquad \rulelabel{rs}\bigfrac{\gamma_1\in
IS^{\star} \;\;\; \gamma_1\tr{\alpha}\gamma_2 }{\gamma_2\in
IS^{\star}}$$
$$\rulelabel{rd}\bigfrac{\gamma\in IS^{\star} \;\;\; \gamma\tr{\alpha_1}\gamma_1
\;\;\; \gamma\tr{\alpha_2}\gamma_2
    \;\;\; \gamma\tr{\alpha_1}\gamma_1 \deriv \gamma\tr{\alpha_2}\gamma_3}{\gamma_3\in
    IS^{\star}}$$}
The rule \rulelabel{rd} adds all the states that are needed to check
redundancy. Indeed, if $\gamma$ can perform both $\tr{\alpha_1}\gamma_1$
and $\tr{\alpha_2}\gamma_2$ s.t. $\gamma\tr{\alpha_1}\gamma_1 \deriv \gamma\tr{\alpha_2}\gamma_3$, 
then $\gamma \tr{\alpha_2}\gamma_2$ would be redundant whenever $\gamma_2 \, \satbis \, \gamma_3$. 

\begin{algorithm}\caption{\texttt{CCP-Partition-Refinement($IS$)}}\label{algo:ccp}
\textbf{Initialization}
\begin{enumerate}
\item Compute $IS^{\star}$ with the rules \rulelabel{is}, \rulelabel{rs} and \rulelabel{rd},
\item $\mathcal{P}^0:=\{IS^{\star}_{d_1}\}\dots \{IS^{\star}_{d_n}\}$,
\end{enumerate}
\textbf{Iteration} $\mathcal{P}^{n+1}:=\mathbf{IR}(\mathcal{P}^n)$

\textbf{Termination} If $\mathcal{P}^n=\mathcal{P}^{n+1}$ then return $\mathcal{P}^n$.
\end{algorithm}

Fig. \ref{fig:part} shows the partitions computed by the algorithm 
with initial states $\pairccp{R'+S}{true}$, $\pairccp{S}{true}$ and $\pairccp{R+S}{true}$.
Note that, as expected, in the final partition $\pairccp{R+S}{true}$ and $\pairccp{S}{true}$ belong to the same block, while
$\pairccp{R'+S}{true}$ belong to a different one (meaning that the former two are saturated bisimilar, while $\pairccp{R'+S}{true}$ is different).
In the initial partition all states with the same store are equated. In $\mathcal{P}^{1}$, the blocks are split by 
considering the outgoing transitions: 
all the final states are distinguished (since they cannot perform any transitions) and 
$\pairccp{T'}{z<5 \sqcup x<5}$ is distinguished from $\pairccp{T}{z<5 \sqcup x<5}$. All the other blocks are not divided, 
since all the transitions with label $x<5$ are redundant in $\mathcal{P}^{0}$ (since $\pairccp{P}{z<5}\mathcal{P}^0 \pairccp{P+Q'}{z<5}$, $\pairccp{P}{z<5}\mathcal{P}^0 \pairccp{P+Q}{z<5}$
and $\pairccp{T'}{z<5 \sqcup x<5} \mathcal{P}^{0} \pairccp{T}{z<5 \sqcup x<5}$). 
Then, in $\mathcal{P}^{2}$, $\pairccp{P+Q'}{z<5}$ is distinguished from $\pairccp{P}{z<5}$ since the transition $\pairccp{P+Q'}{z<5}\tr{x<5}$
is not redundant anymore in $\mathcal{P}^{1}$ (since $\pairccp{T'}{z<5 \sqcup x<5}$ and $\pairccp{T}{z<5 \sqcup x<5}$ belong to different blocks in $\mathcal{P}^{1}$). 
Then in $\mathcal{P}^{3}$, $\pairccp{R'+S}{true}$ is distinguished from $\pairccp{S}{true}$ since the transition $\pairccp{R'+S}{true}\tr{x<5}$ 
is not redundant in $\mathcal{P}^{2}$ (since $\pairccp{P+Q'}{z<5} \not \! \! \mathcal{P}^{2} \pairccp{P}{z<5}$).
Finally, the algorithm computes $\mathcal{P}^{4}$ that is equal to $\mathcal{P}^{3}$ and return it. 
It is interesting to observe that the transition $\pairccp{R+S}{true}\tr{x<5}$ is redundant in all the partitions computed by the algorithm 
(and thus in $\satbis$), while the transition $\pairccp{R'+S}{true}\tr{x<5}$ is considered redundant in $\mathcal{P}^0$ and 
$\mathcal{P}^1$ and not redundant in $\mathcal{P}^2$ and $\mathcal{P}^3$.





\subsection{Termination}\label{sec:term} Note that any iteration splits blocks and never fuse them. For this reason if $IS^{\star}$ is finite, the algorithm terminates in at most $|IS^{\star}|$ iterations.  The proof of the next proposition assumes that $\deriv$ is decidable. However, as we shall prove in the next section, the decidability of $\deriv$ follows from our assumption about the decidability of the ordering relation  $\sqsubseteq$ of the underlying constraint system and Theo. \ref{theo:imp} in the next section.

\begin{proposition}
If $IS^{\star}$ is finite, then the algorithm terminates and the resulting partition coincides with $\satbis$.
\end{proposition}
\begin{proof}
See \cite{sac2012-extended-version}.
\end{proof}

We now prove that if the set $\mathtt{Config}(IS)$ of all configurations reachable from $IS$ (through the LTS generated by the rules in Tab. \ref{tab:labsem}) is finite, then $IS^{\star}$ is finite. 

This condition can be easily guaranteed by imposing some syntactic restrictions on ccp terms, like for instance, by excluding 
either the procedure call or the hiding operator.

\begin{theorem}
If $\mathtt{Config}(IS)$ is finite, then  $IS^{\star}$ is finite.
\end{theorem}
\begin{proof}
See \cite{sac2012-extended-version}.
\end{proof}


\subsection{Complexity of the Implementation}\label{sec:comp}

Here we give  asymptotic bounds for the execution time  of Alg. \ref{algo:ccp}.  We assume that the reader is familiar with the $O(.)$
notation for  asymptotic upper bounds in analysis of algorithms--see \cite{Cormen:09:Book}.  

 Our  implementation of  Alg. \ref{algo:ccp} is  a variant of the original partition refinement algorithm in \cite{Kanellakis:83:PODC} with two main differences: The computation of $IS^{\star}$ according to rules \rulelabel{is}, \rulelabel{rs} and \rulelabel{rd} (line 2, Alg. \ref{algo:ccp}) and the decision procedure for  $\deriv$ (Def. \ref{def:deriv}) needed in the redundancy checks.
 
 Recall that we assume $\sqsubseteq$ to be decidable. Notice that requirement of having some $e$ that satisfies both conditions $(i)$ and $(ii)$  in Def. \ref{def:deriv} suggests that deciding whether two given transitions belong to $\deriv$  may be costly.  The following theorem, however, provides a simpler characterization of  $\deriv$ allowing us to reduce the decision problem of $\deriv$ to that of $\sqsubseteq$. 


\begin{theorem}\label{theo:imp}
$\pairccp{P}{c} \trans{\alpha} \pairccp{P_1}{c'} \deriv \pairccp{P}{c} \trans{\beta} \pairccp{P_1}{c''}$ iff the following conditions hold:
\begin{tabular}{cc}
 (a) $\alpha \sub \beta$ 
& 
(b) $c'' = c' \lub \beta$
\end{tabular}
\begin{proof}
See \cite{sac2012-extended-version}.
\end{proof}
\end{theorem}

Henceforth we shall assume that given a constraint system ${\bf C}$, the function $f_ {\bf C}$ represents  the time complexity of  deciding (whether two given constraints are in) $\sqsubseteq$. The following is a useful corollary of the above theorem. 

\begin{corollary}\label{cor:imp} Given two transitions $t$ and $t'$, deciding whether  $t \deriv t'$ takes $O(f_ {\bf C})$ time.
\end{corollary}

\begin{remark} We introduced $\deriv$ as in Def.  \ref{def:deriv} as natural adaptation of the corresponding notion in  \cite{Bonchi:09:ESOP}. The simpler characterization given by the above theorem is due to particular properties of ccp transitions, in particular monotonicity of the store, and hence it may not hold in a more general scenario. 
\end{remark}

\emph{Complexity.} The size of the set $IS^*$ is central to the complexity of Alg. \ref{algo:ccp} and depends on topology of the underlying transition graph. For  tree-like topologies, a typical representation of many transition graphs, one can show by using a simple combinatorial argument that the size of $IS^*$ is quadratic w.r.t. the size of the set of reachable configurations from $IS$, i.e., $\mathtt{Config}(IS)$. For arbitrary graphs, however, the size of  $IS^*$ may be exponential in the number of transitions between the states of $\mathtt{Config}(IS)$ as shown by the following construction.

\begin{definition}\label{exp}  Let $P^0 = \stopp$ and $P^1 = P.$  Given an even number $n$, define $s_n(n,0)= \stopp, s_n(n,1)=\askp{(true)}{s_n(n,0)}$ and for each $0 \leq i < n \wedge 0 \leq j \leq 1$ let $s_n(i,j)= (\askp{(true)}{s_n(i,j \oplus 1})^{j \oplus 1} \,   
 + \, (\askp{(b_{i,j})}{\stopp}) \, 
+ \, (\askp{(a_i)}{s_n(i+1,j)})$ where $\oplus$ means addition modulo 2. We also assume that (1) for each $i,j:$ $a_{i} \sqsubset b_{i,j}$ and (2)
 for each two different $i$ and $i':$ $a_i \not\sqsubseteq a_{i'},$ and (3) for each two different $(i,j)$ and $(i',j')$:  $b_{i,j}\not\sqsubseteq b_{i',j'}.$
\end{definition}
Let  $IS=\{ s_n(0,0) \}.$  One can verify that the size of $IS^*$ is indeed exponential in the number of transitions between the states of  $\mathtt{Config}(IS).$  

Since Alg. \ref{algo:ccp} computes $IS^*$ the above construction shows that on some inputs Alg. \ref{algo:ccp}  take \emph{at least} exponential time.  We conclude  by stating an upper-bound on the execution time of Alg. \ref{algo:ccp}.
 
\begin{theorem}  Let $n$ be the size of the set of states $\mathtt{Config}(IS)$ and let $m$ be the number of transitions between those states.  Then
$ n \times 2^{O(m)} \times f_{\bf C}$ is an upper bound for the running time of Alg. \ref{algo:ccp}.
\end{theorem}
\begin{proof} 
See \cite{sac2012-extended-version}.
\end{proof}

\section{Concluding Remarks} In this paper we provided an algorithm for verifying (strong) bisimilarity for ccp by building upon the work in \cite{Bonchi:09:ESOP}.  Weak bisimilarity is the variant obtained by ignoring, as much as possible,  silent transitions (transitions labelled with $\true$ in the ccp case).  Neither \cite{Bonchi:09:ESOP} nor the present work deal with this weak variant. We therefore plan to provide an algorithm for this central equivalence in future work.

\bibliographystyle{abbrv}
\bibliography{lab}

\begin{thebibliography}{10}

\bibitem{Aristizabal:11:FOSSACS}
A.~Aristizabal, F.~Bonchi, C.~Palamidessi, L.~Pino, and F.~D. Valencia.
\newblock Deriving labels and bisimilarity for concurrent constraint
  programming.
\newblock In {\em FOSSACS}, pages 138--152, 2011.

\bibitem{sac2012-extended-version}
A.~Aristizabal, F.~Bonchi, L.~Pino, and F.~Valencia.
\newblock Partition refinement for bisimilarity in ccp (extended version).
\newblock Technical report, INRIA-CNRS, 2012.
\newblock Available at:
  \href{http://www.lix.polytechnique.fr/~andresaristi/sac2012.pdf}{http://www.%
lix.polytechnique.fr/\textasciitilde andresaristi/sac2012.pdf}.

\bibitem{Bonchi:09:FOSSACS}
F.~Bonchi, F.~Gadducci, and G.~V. Monreale.
\newblock Reactive systems, barbed semantics, and the mobile ambients.
\newblock In {\em FOSSACS}, pages 272--287, 2009.

\bibitem{Bonchi:06:LICS}
F.~Bonchi, B.~K{\"o}nig, and U.~Montanari.
\newblock Saturated semantics for reactive systems.
\newblock In {\em LICS}, pages 69--80, 2006.

\bibitem{Bonchi:09:ESOP}
F.~Bonchi and U.~Montanari.
\newblock Minimization algorithm for symbolic bisimilarity.
\newblock In {\em ESOP}, pages 267--284, 2009.

\bibitem{Cormen:09:Book}
T.~H. Cormen, C.~E. Leiserson, R.~L. Rivest, and C.~Stein.
\newblock {\em Introduction to Algorithms (3. ed.)}.
\newblock MIT Press, 2009.

\bibitem{Boer:95:TCS}
F.~S. de~Boer, A.~D. Pierro, and C.~Palamidessi.
\newblock Nondeterminism and infinite computations in constraint programming.
\newblock {\em Theor. Comput. Sci.}, 151(1):37--78, 1995.

\bibitem{Fernandez:89:SCP}
J.-C. Fernandez.
\newblock An implementation of an efficient algorithm for bisimulation
  equivalence.
\newblock {\em Sci. Comput. Program.}, 13(1):219--236, 1989.

\bibitem{Ferrari:98:CAV}
G.~Ferrari, S.~Gnesi, U.~Montanari, M.~Pistore, and G.~Ristori.
\newblock Verifying mobile processes in the hal environment.
\newblock In {\em CAV}, pages 511--515, 1998.

\bibitem{Kanellakis:83:PODC}
P.~C. Kanellakis and S.~A. Smolka.
\newblock Ccs expressions, finite state processes, and three problems of
  equivalence.
\newblock In {\em PODC}, pages 228--240, 1983.

\bibitem{Milner:92:ICALP}
R.~Milner and D.~Sangiorgi.
\newblock Barbed bisimulation.
\newblock In {\em ICALP}, pages 685--695, 1992.

\bibitem{Montanari:92:FI}
U.~Montanari and V.~Sassone.
\newblock Dynamic congruence vs. progressing bisimulation for ccs.
\newblock {\em FI}, 16(1):171--199, 1992.

\bibitem{Saraswat:90:POPL}
V.~A. Saraswat and M.~C. Rinard.
\newblock Concurrent constraint programming.
\newblock In {\em POPL}, pages 232--245, 1990.

\bibitem{Saraswat:91:POPL}
V.~A. Saraswat, M.~C. Rinard, and P.~Panangaden.
\newblock Semantic foundations of concurrent constraint programming.
\newblock In {\em POPL}, pages 333--352, 1991.

\bibitem{Victor:94:CAV}
B.~Victor and F.~Moller.
\newblock The mobility workbench - a tool for the pi-calculus.
\newblock In {\em CAV}, pages 428--440, 1994.

\end{thebibliography}
\end{document}